# Phospholipid-Dextran with a Single Coupling Point: a Useful Amphiphile for Functionalization of Nanomaterials

*Andrew P. Goodwin, Scott M. Tabakman, Kevin Welsher, Sarah P. Sherlock, Giuseppe Prencipe, and Hongjie Dai\**

Department of Chemistry, Stanford University, Stanford, California 94305

Email: hdai@stanford.edu

**Received: ;**

*Abstract:* Nanomaterials hold much promise for biological applications, but they require appropriate functionalization to provide biocompatibility in biological environments. For non-covalent functionalization with biocompatible polymers, the polymer must also remain attached to the nanomaterial after removal of its excess to mimic the high dilution conditions of administration *in vivo*. Reported here are the synthesis and utilization singly-substituted conjugates of dextran and a phospholipid (Dextran-DSPE) as stable coatings for nanomaterials. Suspensions of single walled carbon nanotubes were found not only to be stable to phosphate buffered saline (PBS), serum, and a variety of pH's after excess polymer removal, but also provide brighter photoluminescence than carbon nanotubes suspended by poly(ethylene glycol)-

DSPE. In addition, both gold nanoparticles (AuNPs) and gold nanorods (AuNRs) were found to maintain their dispersion and characteristic optical absorbance after transfer into Dextran-DSPE, and were obtained in much better yield than similar suspensions with PEG-phospholipid and commonly used thiol-PEG. These suspensions were also stable to PBS, serum, and a variety of pH's after removal of excess polymer. Dextran-DSPE thus shows great promise as a general surfactant material for the functionalization of a variety of nanomaterials, which could facilitate future biological applications.

**Introduction**

Owing to their unique photophysical and electronic properties, new types of nanomaterials have been implemented in recent years for both *in vitro* and *in vivo* biological applications. Particular attention has been paid to those nanomaterials that absorb light or fluoresce in the near infrared (NIR) range, which boasts substantial penetration into living tissue.[1] Examples include single-walled carbon nanotubes (SWNTs), which exhibit high NIR absorbance[2] and fluorescence[3] and strong Raman scattering signatures.[4,5] SWNTs and other carbonaceous materials[6] have been used as transporters of molecules into cells,[2,7-9] as well as both fluorescent labels[10-12] and contrast agents for NIR-induced cell heating.[2] Gold nanoparticles (AuNPs)[13,14] and nanorods (AuNRs),[15,16] the latter of which have peak absorbances that can reach the NIR range, have been used as biomolecular sensors,[17,18] cell markers,[19] and x-ray contrast agents.[20] Finally, quantum dots (QDs),[21,22] which boast tunable, narrow fluorescence emissions, have been utilized for multiplexed imaging.[23]

Despite the variety of structures and properties of these nanomaterials, to be used *in vivo* they all must satisfy the requirements of biocompatibility including water solubility, suppressed immune response, and lack of toxicity, which is highly dependent on surface functionalization chemistry. In most cases, functionalization is provided by attaching capping ligands or surfactants that also help to prevent aggregation of the nanomaterials and subsequent loss of photophysical properties. While small molecule ligands will not stay bound under high dilution conditions, polymer coatings have been shown to be more stable to changes in aqueous media and can provide more complete coverage of the nanomaterial surface. Thus, an ideal polymer coating for medical applications would: 1) preserve a stable coating even under high dilution conditions, as would be expected *in vivo*; 2) prevent non-specific binding of proteins and macrophages that may result in high background and low blood circulation time in *ex vivo* and *in vivo* applications, respectively, 3) provide accessible functional groups for the augmentation of targeting groups and additional layers of passivation, 4) prevent the rebundling or aggregation of nanomaterials under various conditions and 5) maintain the unique physical properties inherent to the nanomaterial.

For this purpose, a variety of polymeric systems have been developed to form stable protective coatings, by covalent or non-covalent means, around these nanomaterials. In most cases, poly(ethylene glycol) (PEG) is the chief component in many passivating polymeric assemblies, from stealth liposomes[24, 25] to hybrid polymer-carbon nanotubes[2, 7, 26] and quantum dot coatings.[27] Another, less commonly used biocompatible polymer is dextran,[28, 29] a naturally-occurring polymer consisting of α(1→6)-linked glucose monomers that also contains about 5% branching at the 4-position as produced in *Leuconostoc mesenteroides*. The structure of dextran is very interesting for a few reasons, especially in contrast with PEG. First, the hyperbranched structure and packed interior of the dextran results in its globular shape persistence above $M_n$ ~ 10,000,[30] which makes it similar to other manmade hyperbranched structures like dendrimers[31-35] and hyperbranched polyols.[36, 37] This, in turn, may allow a coating more resistant to protein adhesion than a linear polymer. As hyperbranched structures often require multiple steps or precise synthetic conditions to obtain useful polymer scaffolds, a naturally occurring analog may save much time and effort towards the synthesis of novel hyperbranched materials. Second, its polyhydroxylated structure makes dextran a very hydrophilic polymer, which may provide a sharp contrast with any hydrophobic components and thus provide facile self-assembly. Finally, dextran contains many accessible hydroxyl groups, which may be functionalized further with short PEG units to obtain hyperbranched polymeric structures that show improved circulation *in vivo* when compared to linear analogs.[26, 38] In addition, these sites allow significant loading of targeting groups like the RGD peptide motif[39] or mannose.[40] This same property, however, often precludes the use of dextran in structure-specific applications because of the difficulties in attaching specific numbers of functional groups to the dextran.[41-43]

In this paper, we report the synthesis of a phospholipid-dextran conjugate, in which the phospholipid, 1,2-distearoyl-*sn*-glycero-3-phosphoethanolamine (DSPE), is bound at a single point to the reducing end of the dextran. We found that not only was this material able to form self-assemblies easily in water, but also formed stable coatings on carbon nanotubes, gold nanoparticles, and gold nanorods. These dextran-nanomaterial coatings showed stability to a range of pH's, salt conditions, and introduction of serum. Finally, the dextran-coated materials showed improved photophysical properties when compared with similar suspensions in PEGylated-phospholipid and commonly used thiol-terminated PEG (PEG-SH). Specifically, SWNTs coated with phospholipid-dextran showed a several-fold enhancement in photoluminescence, which could facilitate NIR imaging applications of nanotubes, and phospholipid-dextran was able to suspend AuNRs, which could not be achieved with PEGylated-phospholipid. Thus, phospholipid-dextran conjugates have boasted versatile and high quality suspensions that to our knowledge have been previously unobserved in the adaptation of nanomaterials for biological applications.

**Materials and Methods**

**General Materials and Methods.** Fluorescence emission and excitation spectra were obtained using an ISA/SPEX Fluorolog 3 equipped with a 450 W Xe lamp, double excitation and double emission monochromators and a digital photon-counting photomultiplier. Slit widths were set to 5.0 nm band-pass on both excitation and emission monochromators. UV-Vis spectra were taken on a Cary 6000i spectrophotometer. Carbon nanotube spectra were taken in double beam experiment using two matching 1-mm pathlength glass cuvettes. Nanoparticle and nanorods were measured in a 1-cm path-length quartz cuvette in a single beam experiment. $^1$H Nuclear Magnetic Resonance spectra were obtained on an Oxford AS400 at 400 MHz. NMR chemical shifts are reported in ppm and calibrated against DMSO-$d_6$ (δ 2.49). Dynamic Light Scattering measurements were obtained on a Brookhaven Instruments 90Plus Particle Size Analyzer. The polydispersity reported is the ratio of the statistical variance of the sample to the square of the mean diameter, as calculated by the instrument software. Atomic Force Microscopy (AFM) was conducted on a Multimode Nanoscope IIIa AFM system (Veeco).

Photoluminescence excitation (PLE) measurements in the NIR were performed utilizing a home-built setup. A short arc lamp (Osram XBO 75W/2 OFR 75W Xenon lamp installed into Oriel 66907 Arc Lamp Source) and a monochromator (Oriel 7400 Cornerstone 130 monochromator) were used to supply excitation light in the 550 nm – 840 nm range in 10 nm steps. The excitation light was focused onto a sample placed in a 1 mm path quartz cuvette. The room temperature sample photoluminescence was collected at the opposite cuvette wall and the PL spectrum was recorded using a second monochromator (Acton SpectraPro 2300i) and a liquid nitrogen-cooled InGaAs array detector (Princeton Instruments OMA V 1024-2.2 LN) in the 900–1500 nm range. As-obtained PL spectra were scaled by the measured excitation power (measured using Oriel 71580 calibrated Si photodiode) before obtaining a PLE spectrum by interpolating the thirty measured PL spectra. The bandpass used for emission and excitation was 15nm.

**Synthesis of Dextran-DSPE (1)**. In a scintillation vial with stirbar, 200 mg dextran (0.0114 mmol, $M_n$ ~ 17,500; MP Biomedicals, Inc.) and 32 mg N-hydroxysuccinimide (NHS, 0.286 mmol, Pierce) were dissolved in 8 mL dry DMSO (Fluka). This solution was kept at 60°C until all solid dissolved, then cooled to RT. 122 mg N,N'-dicyclohexylcarbodiimide (DCC, 0.586 mmol, Aldrich) was added and the mixture agitated until the DCC was observed to dissolve. In a separate vial, 88 mg 1,2-distearoyl-*sn*-glycero-3-phosphoethanolamine (DSPE, 0.117 mmol, Lipoid LLC) and 34 μL triethylamine (0.228 mmol, EMD) were combined in 2 mL chloroform (EMD) and sonicated briefly until any clumps of solid were dispersed in the solvent. This suspension was added to the first solution carefully to avoid spilling along the sides. The combined suspension was stirred at 60°C for 2 d, then cooled to RT. The suspension was then subjected to an air stream until the presence of chloroform could no longer be observed by smell. A mixture of the remaining suspension and 24 mL distilled water were transferred to a centrifuge tube, centrifuged at 10,000 *g* for 30 min, and coarsely filtered. The filtrate was transferred to a 3500 MWCO regenerated cellulose dialysis tube (Fisher) and dialyzed 3 d against several changes of distilled water. Finally, the dialysate was centrifuged at 10,000 *g* for 30 min, refiltered through a 200 nm pore-size filter to remove any remaining solid, and freeze-dried to obtain a fluffy, white powder (170 mg, 81%). This final purification was repeated prior to characterization. $^1$H NMR (DMSO-$d_6$): δ 0.84-0.90 (m, 6H), 1.21-1.32 (m, 32H), 1.49-1.53 (m,

4H), 2.25-2.30 (m, 6H), 3.05-3.87 (m, ~550H), 4.46-4.62 (m, 108H), 4.62-4.78 (m, 108H), 4.78-5.02 (m, 216H).

**Determination of Critical Micelle Concentration.** 10 μL of a stock solution containing 0.10 mg/mL pyrene in chloroform was added to each of eight vials. These vials were placed under vacuum and allowed to dry for at least 30 min. Solutions of **1** in phosphate buffered saline (PBS) were made at concentrations indicated in the text, and 3 mL of the appropriate concentration was added to each vial. The vials were agitated gently overnight to allow pyrene uptake. The fluorescence emission of pyrene ($\lambda_{exc}$ = 340 nm) was measured for each solution. After subtracting a background of any PBS fluorescence, the ratio of the emission at 373 nm to the emission at 383 nm was determined and plotted against the log of the concentration of **1**. The inflection point as determined by the best fit of the data above and below the CMC gave a CMC of 1.8 μM.

**Suspension of SWNTs.** Excellent SWNT suspensions were obtained via the following steps. First, a 3.6 mg/mL (200 μM) solution of dextran-17-DSPE (**1**) in deionized ultrafiltered water was prepared by adding **1** to the water, sonicating for 5 min, then centrifuging the mixture for 10 min at 25,000 *g*. The supernatant was then used for suspension. The rest of this method of suspending nanotubes has been described previously.[2] Briefly, 0.75-2.0 mg raw HiPCO SWNTs (Unidym, Inc. >65% carbon) were placed in a 20 mL scintillation vial, followed by 4 mL of the above solution for every 1.0 mg of NTs in the vial. This mixture was vortexed for a few min until almost all of the NTs were below the solvent line, then sonicated 30 min. The sonicator water bath was replaced with fresh RT distilled water and the suspension was sonicated an additional 30 min. The dark suspension was centrifuged for 6 h at *ca.* 25,000 *g*, and the supernatant was recovered carefully via pipette for subsequent experiments. The excess surfactant was removed through vacuum filtration through a 0.1 μm polycarbonate membrane (Millipore) and extensive washing with deionized ultrafiltered water. The best results were obtained by not allowing the filter to dry, although most SWNTs that adhered to the filter could be resuspended through brief (< 5 min) vortexing or bath sonication of the filter in water. For comparison to solutions obtained prior to excess polymer removal, the filtered solution was concentrated to the original sample volume via a 100 kDa centrifuge filter (Millipore).

Suspensions of SWNTs in mPEG(5000)-DSPE (Laysan Bio) were prepared as described in the literature.[2]

**Suspensions of Gold Nanoparticles.** 5 mg of **1**, mPEG(5000)-DSPE, or PEG(5000)-SH (Nektar) were dissolved in 5 mL of a citrate-stabilized 20 nm gold colloid solution (Sigma). The mixture was sonicated 15 min, then transferred to a 3500 MWCO regenerated cellulose tubing a dialyzed overnight with several changes of distilled water. The resultant solution was then centrifuged at 25,000 $g$ for 10 min, which caused the NPs to form a pellet. The supernatant was removed and fresh water was added. This washing procedure was performed four times total to remove excess polymer, after which fresh water was added to bring the final volume to 5 mL.

**Gold Nanoparticle Suspension Stability Tests.** For PBS stability tests, 450 μL of the Au NP solution was combined with 50 μL 10X PBS (Gibco). For pH tests, 450 μL of the solution was combined with 50 μL of an appropriate solution of HCl or NaOH. For serum tests, 450 μL solution, 50 μL 10X PBS, and 500 μL Fetal Bovine Serum (Gibco). All test solutions were agitated overnight prior to UV-Vis measurement and photographing.

**Synthesis of Gold Nanorods.** Gold nanorods were synthesized via a literature procedure by Nikoobakht and El-Sayed.[16] For the experiments described in the main text, the approximate size of the nanorods is 35x15x15 nm as determined by matching the obtained absorption spectrum (Figure 5A) with those in their report. Immediately prior to use, the cetyl trimethylammonium bromide (CTAB)-stabilized NR suspensions were sonicated at least 10 min to redissolve any precipitated CTAB.

**Suspension of Gold Nanorods.** To 4 mL CTAB-stabilized NRs suspension, 1 mL N-methylformamide (NMF, Aldrich) was added. The suspension was mixed and then centrifuged at 25,000 *g* for 10 min to remove CTAB. The supernatant was removed and replaced with a fresh 4:1 water:NMF mixture, vortexed to resuspend the pelleted NRs, and recentrifuged. The supernatant was removed, 8 mL water was added to the pelleted NRs. The mixture was vortexed back into solution, and 8 mg of either **1** or mPEG(5000)-DSPE was dissolved. The suspension was sonicated 15 min, then transferred to a 3500 MWCO regenerated cellulose tubing and dialyzed overnight with several changes of distilled water. The resultant solution was then centrifuged at 25,000 *g* for 10 min, which caused the NRs to pellet. The supernatant was removed and fresh water was added. This washing procedure was performed four times total to remove excess polymer surfactant, after which the final volume was 5 mL.

**Gold Nanorod Stability Tests.** Stability tests were performed as with the NPs described above.

**Results and Discussion**

**Synthesis of Dextran-DSPE.** Early in our work, we found that dextran alone was unable to provide a stable suspension of nanotubes.[29] Studies in our labs have shown that PEG conjugates of 1,2-distearoyl-*sn*-glycero-3-phosphoethanolamine (DSPE) are able suspend carbon nanotubes well to prevent aggregation under serum conditions with excess polymer removed.[2, 26] In addition, phospholipids also boast natural occurrence, biocompatibility, and strongly hydrophobic components, all of which provide advantages over other nanotube binding groups such as pyrene[44] or porphyrins.[45] Thus, the challenge was to join the DSPE and dextran together.

To mimic the PEG-DSPE system, we decided to attach DSPE to the dextran at a single point only. This can be done at the reducing, or anomeric, end of the dextran, which differs in reactivity from the hydroxyl groups on the rest of the polymer.[46] Previous syntheses that utilized the reducing end of dextran first required the coupling of a bifunctional amine by reductive amination, followed by the attachment of a molecule of interest.[46, 47] However, in many cases the yield of the reductive amination is limited,[46] and it would be very difficult to separate free dextran from a phospholipid conjugate. We reasoned that a one-step ligation using a different method would not only decrease the number of steps involved in the synthesis but also improve the efficiency of purification of the final dextran conjugates.

In common syntheses of glycosylamines, the alkyl and sugar components are joined simply by acid-catalyzed condensation of the amine to form a stable cyclic aminal.[48] Interestingly, we could find no literature reports for this reaction having been performed with higher molecular weight dextran. We surmised that given the higher molecular weight of the dextran, the reaction would require stronger conditions. However, highly elevated temperatures resulted in a yellowing of the DSPE and low yield, which has been attributed to a thermally-promoted rearrangement special to phosphoethanolamines.[49]

Instead, a chemically-promoted dehydration was tried with an excess of N-N'-dicyclohexylcarbodiimide being added as a dehydration agent, along with N-hydroxysuccinimide both to facilitate the DCC condensation and as a weak acid (Scheme 1). A similar strategy, albeit with CuCl as a catalyst, has been used to join phenols to oligosaccharides.[50] Gratifyingly, after stirring at 60°C for 2 d, the DSPE was found to add to the dextran in a 1:1 ratio by $^1$H

NMR. A 4:1 cosolvent mixture of DMSO and chloroform was required to both solubilize the dextran and suspend the DSPE to allow the reaction to occur.

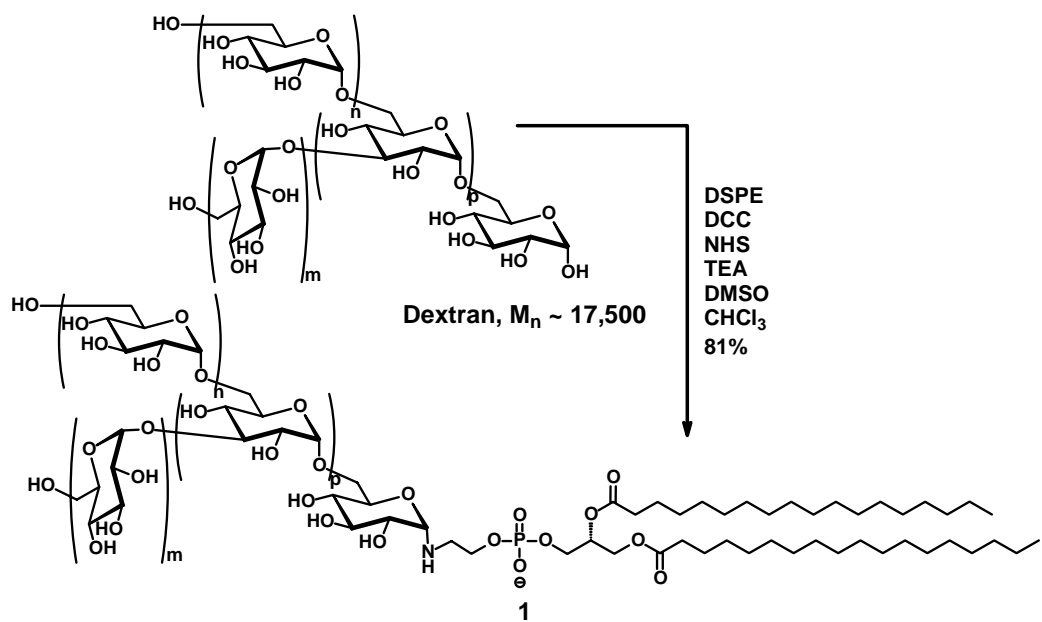

**Scheme 1:** Synthesis of dextran-DSPE at the reducing end of dextran.

To our knowledge, this is the first time that a phospholipid has been conjugated by a single-point reaction with dextran. First, it should be stated that the coupling cannot be as a result of DMSO-induced Moffatt oxidation, since the reaction also worked in a DMF:toluene mixture. Second, in test trials it was found that a reaction of dextran and DSPE with all of the above reagents except for DCC the reaction proceeded very slowly. Without NHS, the reaction did not proceed at all. From these observations, we propose that this reaction may be operating under two competing mechanisms. By itself, the NHS may function as a weak acid and promote the aminalation of the DSPE to a certain extent, which has been observed in acid-catalyzed synthesis of glycosylamines.[48] Other procedures have reported the synthesis of O-glucopyranosyl-N-hydroxysuccimides that were based on Lewis Acid assisted coupling of NHS to the anomeric portion of glucose,[51] which were then reacted with amines after addition of EDC.[52] In this case, the NHS acted as a bifunctional linkage between the glucose and the amine. Thus the dominant reaction pathway may proceed through a DCC-induced dehydration with NHS stabilizing the active intermediate, followed by amine attack to form the aminal. While more studies are certainly warranted to conclude a mechanistic pathway, it should be noted that this reaction is difficult to analyze owing to the complicated structures and solubility requirements of its reactants.

**Self-assembly of 1.** An interesting property of each of the components of this product is their selective solubility in most solvents: dextran is only soluble in water and very strongly polar aprotic solvents, while phospholipids are more soluble in fairly non-polar media and tend to self-assemble in more polar solvents. Thus, in aqueous conditions hydrophobicity-driven self-assembly would be expected. To determine the presence of assemblies, varying amounts of **1** were dissolved in PBS and incubated with pyrene overnight. As more pyrene is encapsulated in a hydrophobic environment, the fine spectra of the pyrene unimer change, resulting in a change in the peak heights of the emission spectra.[53, 54] A plot of the ratios of fluorescence emission at 373 nm to 383 nm vs. log of concentration (Figure 1A) shows a sharp change around 1.8 μM, indicating the Critical Micelle Concentration (CMC), a value is very similar to that of PEG-DSPE.[55] It should be noted that at higher concentrations of **1**, a very small amount of insoluble solid was present. We suspected either that this was excess DSPE, which would interfere with the assembly, or that the concentration of **1** had become large enough to induce the formation of other superstructures. However, simply centrifuging away the solid was sufficient to obtain solutions appropriate for Dynamic Light Scattering, which showed the presence of assemblies that were 70 nm in diameter (Figure 1B). Given that the average hydrodynamic diameter of each dextran molecule is *ca.* 6-7 nm[56] and the length of the phospholipid is approximately 3-3.5 nm, this would indicate at least a partial bilayer to account for the added size of the assembly.

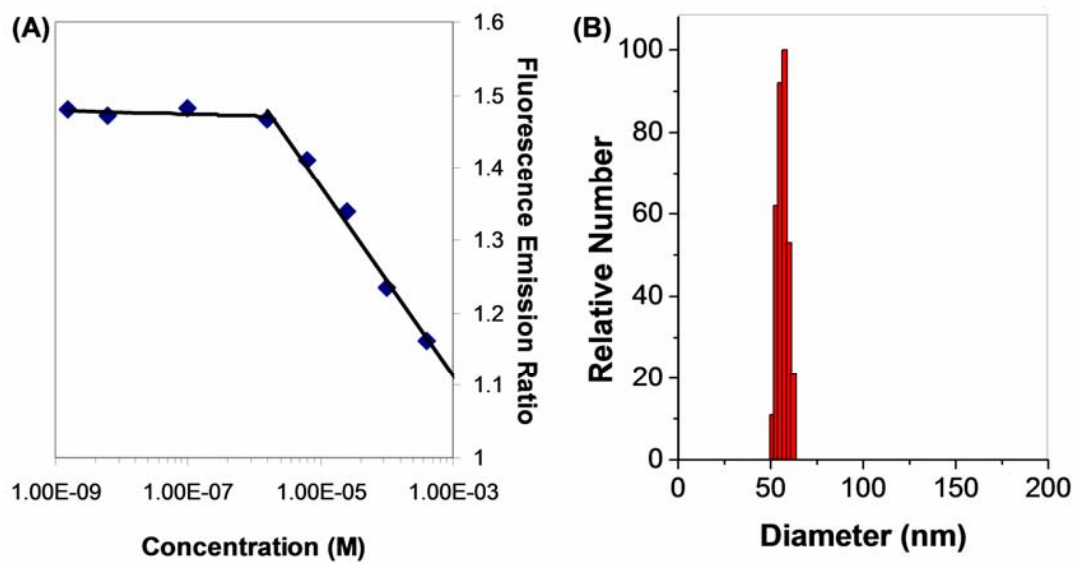

**Figure 1:** (A) Ratio of fluorescence emission of pyrene at 373 nm to 383 nm ($\lambda_{exc}$ = 340 nm) vs. log of concentration of dextran-DSPE (spectra found in Figure S2). This ratio indicates a change in pyrene monomer vibronic band intensities that correlates with a change in pyrene solvent environment, or in this case micelle formation. The inflection point in the curve indicates a Critical Micelle Concentration of *ca.* 1.8 µM. (B) Dynamic Light Scattering of 200 µM **1** in water shows assemblies of 70 nm in diameter, with an average particle polydispersity of 0.100.

**Functionalization of Carbon Nanotubes by 1.** Carbon nanotubes are especially challenging materials to suspend, because they require non-covalent functionalization to maintain their important spectroscopy properties, such as NIR absorption features,[2] photoluminescence,[3] and Raman scattering.[4] Our dextran-DSPE proved to be useful amphiphiles for nanotube suspension. To make dextran-DSPE-suspended SWNTs, dry, raw HiPCO-produced SWNTs were sonicated for 1 h in the presence of a predissolved, precentrifuged 200 µM solution of **1** in water, followed by removal of unsuspended aggregates by centrifugation. The UV-Vis spectrum of this material is shown in Figure 2A, and depicts a typical HiPCO-produced carbon nanotube UV-Vis spectrum.[3]

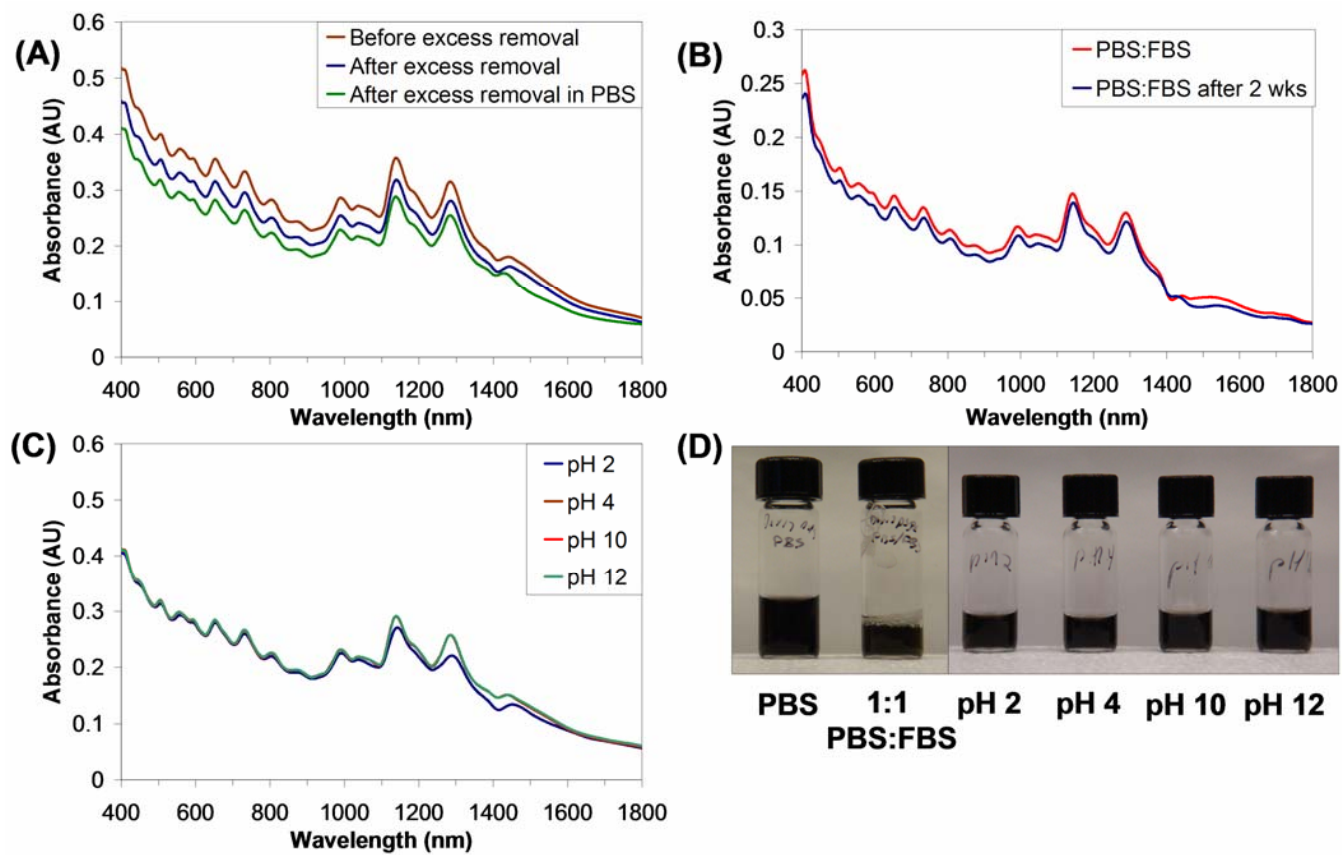

**Figure 2:** UV-Vis-IR absorption spectra of SWNTs encapsulated by 1 after 6 h centrifugation at 25,000 g (orange) and after removal of excess 1 by vacuum filtration (blue), and after transfer of this solution into PBS. (B) To demonstrate stability of suspension to serum, the suspension was transferred to a 1:1 mixture of PBS:FBS (red). Despite two weeks of incubation, the suspension yield did not decrease significantly (blue). (C) Exposure to different pH conditions led only to a decrease in suspension yield for pH 2, while the other suspensions gave exactly the same spectrum. (D) Photograph of SWNT suspensions depicted in (A) and (C) as indicated. The black color is typical of good SWNT suspensions.

While suspensions of carbon nanotubes can be obtained with a variety of different surfactants,[57] few are both non-toxic and able to withstand a variety of different conditions without excess surfactant present. In our case, removal of excess **1** by vacuum filtration through a 100nm-pore polycarbonate filter gave a spectrum very similar to that obtained with excess intact (Figure 2A), and the completeness of excess removal was confirmed by the absence of free Dextran-DSPE in images obtained by AFM (Figures S2 and S3), which in poorly washed samples would be visible in aggregates of about 100 nm. Moreover, the nanotube suspensions were stable without aggregation in PBS and other biological buffer solutions. As other groups have done,[58] we ascribe the ability of **1** to suspend SWNT to the formation of hydrophobic zones created in the self-assembly of **1**, followed by the sequestering of the SWNTs in that assembly.

Following these initial promising studies, the DSPE-dextran-nanotube conjugates were subjected to more stringent tests. First, the PBS suspension was diluted halfway with Fetal Bovine Serum (FBS) to mimic conditions in blood circulation. The UV-Vis spectra in Figure 2B show that the nanotubes were stable to these conditions both after 24 h and two weeks. The carbon nanotube suspensions were also subjected a variety of different acid and base conditions to test the stability after overnight incubation. At pH 4, pH 10, and pH 12, the carbon nanotubes showed no evidence of bundling or precipitation, as the UV-Vis spectra of each of these all coincide (Figure 2C). At pH 2, there was a slight loss in signal. This may be due to degradation of dextran under acidic conditions, either due to hydrolysis of the glycosidic linkages within dextran or hydrolysis of the dextran-DSPE bond.

The photoluminescence of the suspensions of nanotubes were also quite favorable. It can be difficult to maximize the photoluminescence of carbon nanotubes, which depends on three factors: 1) the debundling of the SWNTs, 2) the number of defects on the SWNTs, and 3) the type of stabilizing agent used to create the suspensions. Shown in Figure 3A are photoluminescence emission (PLE) spectra of a washed, Dextran-DSPE suspension of SWNTs in PBS. Peaks can be seen corresponding to (7,5), (7,6), (8,4), and (9,4), which is typical for a HiPCO-produced sample.[59] Interestingly, this sample proved to have a significantly better overall photoluminescence than a similarly prepared sample of PEG-DSPE-produced SWNTs (Figure 3B). There are a few possible reasons for this. First, the persistent size of the dextran may prevent the SWNTs from rebundling better than PEG-DSPE, leading to more single tubes in solution. Second, use of dextran may have provided a milder sonication environment that leads to fewer defect sites on the tubes, or may have left longer tubes in solution. However, further investigations are needed to elucidate these factors.

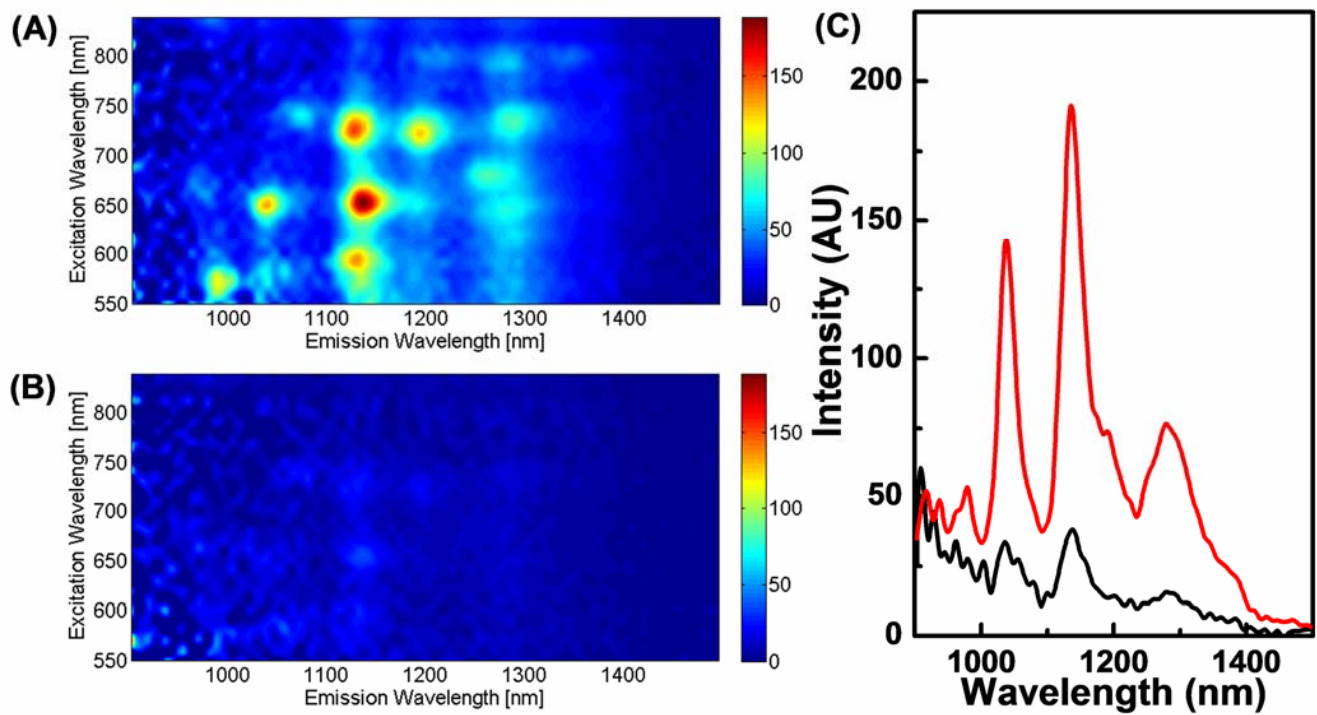

**Figure 3:** Photoluminescence excitation spectra of SWNTs (normalized to tube concentrations measured by UV-Vis) suspended in (A) **1** and (B) mPEG(5000)-DSPE after removal of excess

polymer. (C) Fluorescence emission spectra ($\lambda_{exc}$ = 650 nm) of SWNTs in **1** (red) and mPEG(5000)-DSPE (black).

One of the applications of stable, biocompatible suspensions of carbon nanotubes is their use as fluorescence tags for both *ex vivo* and *in vivo* imaging. SWNTs contain both excitation and emission wavelengths within the tissue transparency window, which corresponds to an area of the visible and NIR spectrum to which living tissue has increased transparency.[1] Shown in Figure 3C is the fluorescence emission at 650 nm excitation. The overall fluorescence emission of the **1**-suspended tubes shows approximately a four-fold increase in total fluorescence emission over the PEG-DSPE-suspended tubes.[10, 26, 60] This, in practice, would correspond to a fourfold increase in sensitivity, which would be beneficial for both cell[10] and *in vivo* imaging.

**Stable Functionalization and Suspension of Gold Nanoparticles and Nanorods.** The most common method of gold nanoparticle functionalization is through a gold-thiol bond, but both sodium dodecylsulfate (SDS)[61] and Triton-X-100,[62] a PEG-based amphiphile, have previously been used to template the synthesis of and stabilize gold nanoparticles. Our DSPE-dextran was shown to be a suitable coating for gold nanomaterials as well. This was shown initially with spherical gold nanoparticles (AuNPs), which are commercially available. First, a citrate-stabilized gold colloid solution was combined with **1** and then sonicated to displace the citrate. Immediately, a slight color change could be observed, which we surmised was a change in plasmon resonance as a result of a change in surface coating and thus the dielectric environment of the AuNP.[63-65] The excess citrate was dialyzed away to ensure full assembly of the polymer on the nanoparticles, and then excess polymer was removed via a series of centrifuge washings to obtain a stable, washed suspension of gold nanoparticles. In contrast, mPEG(5000)-DSPE gave stable suspensions of gold NP, but only at about 90% of the yield found with **1**, and AuNPs coated with PEG-SH after removal of excess PEG-SH showed aggregation of Au NPs (Figure 4A).

These particles also showed excellent stability in other aqueous media. As shown in Figure 4B, dilution into PBS did not change the shape of the UV-Vis trace, but under the same conditions citrate-stabilized particles aggregate and precipitate from solution. The DSPE-dextran coated AuNPs were also stable to serum (Figure 4B) with characteristic optical absorbance largely intact. As with the carbon nanotube suspensions, the AuNP suspensions also showed good stability towards other pH's, with the greatest stability at pH's 4 and 10 and the least stability at pH 2 (Figure 4C). However, it should be noted that part of this decrease in absorbance may be due to a change in the surface plasmon due to the changing salinity of the media with higher acid or base addition.

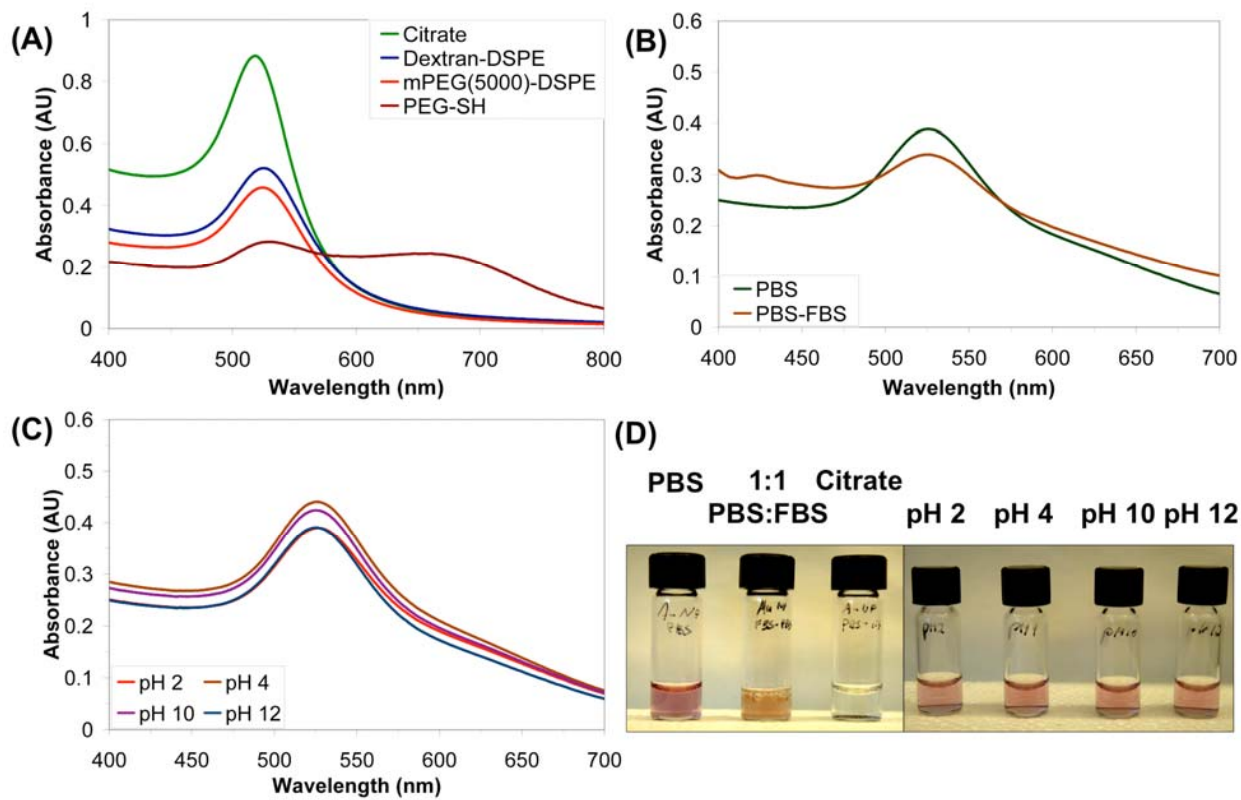

**Figure 4:** (A) UV-Vis absorption spectra of 20 nm AuNPs suspended by 1 (blue), mPEG(5000)-DSPE (red), and PEG-thiol (brown) in water. Excess polymer has been washed away four times for 1 and mPEG(5000)-DSPE, while the PEG-thiol has only been washed twice. The large peak at 660 nm for PEG-thiol is due to NP aggregates in the suspension formed after removal of excess PEG-thiol from the solution. (B) UV-Vis spectra of 20 nm AuNPs suspended by 1 in PBS (green) and 1:1 PBS:FBS (brown). The small peak at 425 nm is due to the large absorbance of FBS in this region. (C) UV-Vis spectra of 20 nm AuNPs suspended by **1** after excess polymer removal at pH 2 (green), 4 (brown), 10 (purple) and 12 (red). (D) Photographs of solutions depicted in (B) and (C). The orange color in the PBS:FBS solution comes from the pink AuNPs and the yellow serum, and the aggregated citrate-suspended particles can be seen at the bottom of their vial.

Next, DSPE-dextran functionalization was attempted with cylindrical gold nanorods.[16] AuNRs are of particular interest because their longitudinal surface plasmon resonances can be tuned to fall in the NIR range. These suspensions were made in a similar manner to the gold nanoparticles, followed by removal of excess **1**. While there was a decrease in nanorod yield as a result of the transfer, the peaks remained sharp (Figure 5A). There was also a slight red shift in the longitudinal plasmon peak at 636 nm and blue shift in longitudinal peak at 515 nm, presumably due to the change in environment around the nanorod. Also, a shoulder appears at around 750 nm, which is most likely due to slight aggregation of the rods.[64] CTAB-encapsulated NRs, however, showed a dramatic decrease in absorbance and loss of color after washing away excess surfactant. Thus functionalization by **1** stabilizes the AuNRs against precipitation. The NRs also remained similarly suspended after transfer into PBS solutions and serum (1:1 PBS:FBS) (Figure 5B). While there is a slight red shift of the longitudinal plasmon peak and blue shift of the transverse peak, the overall intensity remains fairly constant. Finally, as before, the DSPE-Dextran-suspended NRs were tested for their stability to different pHs. As with the NTs, the best suspensions were at pH 4, 10, and 12, while the NRs incubated at pH 2 showed a severe loss of color. We speculate that again the loss of signal is due to dextran instability at very acidic pH.

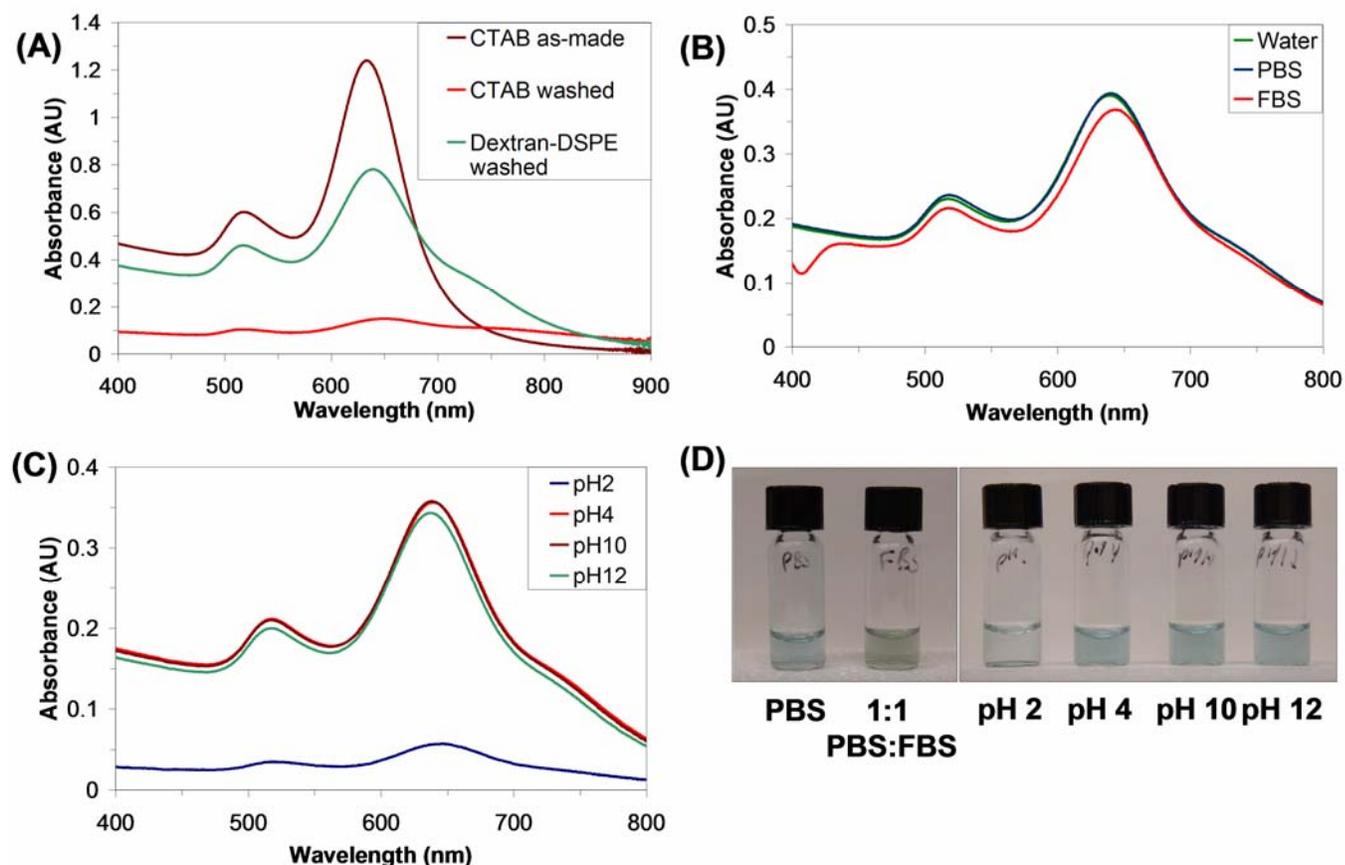

**Figure 5:** (A) UV-Vis-IR absorption spectra of AuNRs suspended by CTAB as-made (brown), after removal of CTAB (red), and suspended by **1** after excess polymer removal (green). (B) UV-Vis spectra of AuNRs suspended by **1** after excess polymer removal in PBS (green) and 1:1 PBS:FBS (red). (C) UV-Vis spectra of AuNRs suspended by **1** after excess polymer removal at pH 2 (blue), 4 (brown), 10 (red) and 12 (green). (D) Photographs of solutions depicted in (B)

and (C) as indicated. The 1:1 PBS:FBS solution is green due to the blue color from the AuNRs and the yellow color from the serum.

Since Dextran-DSPE self-assembles readily in aqueous solution but lacks a functional group that binds specifically to the surface of a nanomaterial, most likely the nanomaterials were encapsulated within the Dextran-DSPE assembly. As with Triton-X and SDS, it appears that the Dextran-DSPE encapsulated the gold nanomaterials via hydrophobic interactions with the metal surface, while attempts to suspend AuNRs with PEG-SH resulted in a complete loss of UV-Vis absorbance after excess PEG-SH removal, most likely due to aggregation (unpublished data). Thus, this material may have implications for alternate forms of gold nanomaterial suspension beyond gold-thiol bonds. However, AuNRs could not be suspended by mPEG(5000)-DSPE either (Figure S4), which complicates an explanation based simply on micelle formation. We speculate that the reason that **1** works better is that the structure of the hyperbranched dextran is able to prevent the reassociation of the nanorods more effectively than the linear PEG, providing a more stable coating. As discussed in the introduction, the crowded interior of dextran provides a shape persistence not generally found in linear polymers such as PEG, and this in turn can provide a shield against the stacking of other nanomaterials, resisting aggregation. In addition, the bulk of the dextran head group may provide better coverage on the surface of the metal, again prevent reassociation. Finally, it may be that by its assembly the Dextran-DSPE is better able to encapsulate cylindrical materials than PEG derivatives.

**Conclusion**

A singly-bonded conjugate of dextran-17 and DSPE was synthesized and shown to be an excellent surfactant for non-covalent functionalization of several nanomaterials for potential biological applications. A single molecule of phospholipid was attached by a synthetic route that utilized only the reducing end of dextran, and the amphiphile was found to exhibit a low CMC of 1.8 μM. Dextran-DSPE was also found to provide PBS- and serum-stable suspensions of carbon nanotubes that were more fluorescent than those with mPEG(5000)-DSPE. Finally, the Dextran-DSPE was able to suspend both spherical gold nanoparticles and cylindrical gold nanorods with similar stability and with better encapsulation yield than mPEG(5000)-DSPE, most likely due to the ability of dextran to cover large areas of the nanomaterial and protect the nanomaterial from reaggregating. Having established Dextran-DSPE as a safe and stable coating of nanomaterials, we plan to implement this material for *in vivo* imaging and therapy applications.

**Acknowledgements.** The authors would like to thank the Center on Polymer Interfaces and Macromolecular Assemblies (CPIMA) for use of their Dynamic Light Scattering Instrument. This work was funded by NIH Grants 1U54-CA119367 (Center for Cancer Nanotechnology Excellence – Therapeutic Response) and R01-CA135109.

**Supporting Information Available.** NMR spectra of **1**, AFM images of **1**-suspended SWNTs. This material is available free of charge via the Internet at http://pubs.acs.org.

TABLE OF CONTENTS

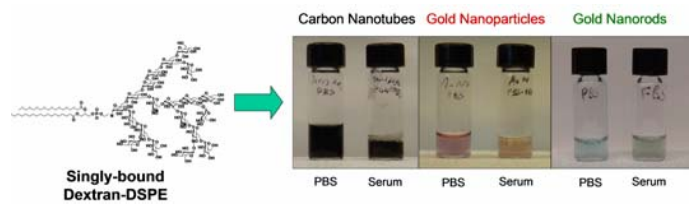